\documentclass[%
superscriptaddress,
 preprint,
 amsmath,amssymb,
]{revtex4-1}

\usepackage{graphicx}
\usepackage{dcolumn}
\usepackage{bm}
\usepackage{xcolor}

\begin{document}

\title{Spectroscopy of fractional orbital angular momentum states}

\author{B. Berger}
\author{M. Kahlert}
\author{D. Schmidt}
\author{M. A{\ss}mann}

\affiliation{Experimentelle Physik 2, Technische Universit\"at Dortmund, 44227 Dortmund, Germany
}

\date{\today}

\begin{abstract}
We present an approach for measuring the orbital angular momentum (OAM) of light tailored towards applications in spectroscopy and non-integer OAM values. It is based on the OAM sorting method (Berkhout et al., Phys. Rev. Lett. {\bfseries 105}, 153601 (2010)). We demonstrate that mixed OAM states and fractional OAM states can be identified using moments of the sorted output intensity distribution and OAM states with integer and non-integer topological charge can be clearly distinguished. Furthermore the difference between intrinsic OAM and total OAM for fractional OAM states is highlighted and the importance of the orientation of the fractional OAM beam is shown. All experimental results show good agreement with simulations. Finally we discuss possible applications of this method for spectroscopy of semiconductor systems such as exciton-polaritons in microcavities.
 \end{abstract}
\maketitle

\section{Introduction}

Light exhibits several momentum degrees of freedom. Besides the linear momentum $p=\hbar k$ and the spin angular momentum (SAM) corresponding to circular polarization $\sigma^{+}$ and $\sigma^{-}$, a light beam can additionally carry orbital angular momentum (OAM) of $l\hbar$ per photon, a concept which was introduced by Allen et al. in 1992 \cite{Allen_1992}. A light beam with OAM has helically shaped phase fronts and can mathematically be described by a phase cross section of $\exp{(il\phi)}$, where $l$ denotes the topological charge as an indicator value for the respective OAM state \cite{Allen_1992,Molina-Terriza_2007,Franke_Arnold_2008} and $\phi$ is the azimuthal angle. Such light fields show a helical phase gradient of $2\pi l$. Compared to the SAM of light with only two orthogonal states, $l$ can therefore take any positive or negative value. One can therefore identify two different classes of OAM states: states with integer or fractional values of $l$. Integer valued $l$ form an orthogonal set of states and due to the unlimited range of $l$, in principle any amount of additional information can be stored in the OAM degree of freedom of the light beam \cite{Molina-Terriza_2001, Gibson_2004, Leach_2010}. In recent years, it has been demonstrated that a set of orthogonal integer OAM states provides a means for secure free-space optical data communication processes \cite{Berkhout_2010, Gibson_2004, Wang_2012}. Further, many possible applications of light beams carrying OAM have been realized in diverse fields, such as optical tweezers \cite{He_1995, Padgett_2011}, quantum and nano optics \cite{Toyoda2012, Bliokh_2015}. For these states, the helical phase gradient is smooth everywhere. For states with non-integer values of $l$, the total helical phase gradient is not an integer multiple of $2\pi$ anymore and the phase gradient must necessarily show at least one abrupt step. The direction of this step defines a preferred direction. The complex phase structure of these states renders them unstable upon propagation \cite{Gotte_2008}.

\begin{figure}[h!]
\centering
	\includegraphics[width=0.75\columnwidth]{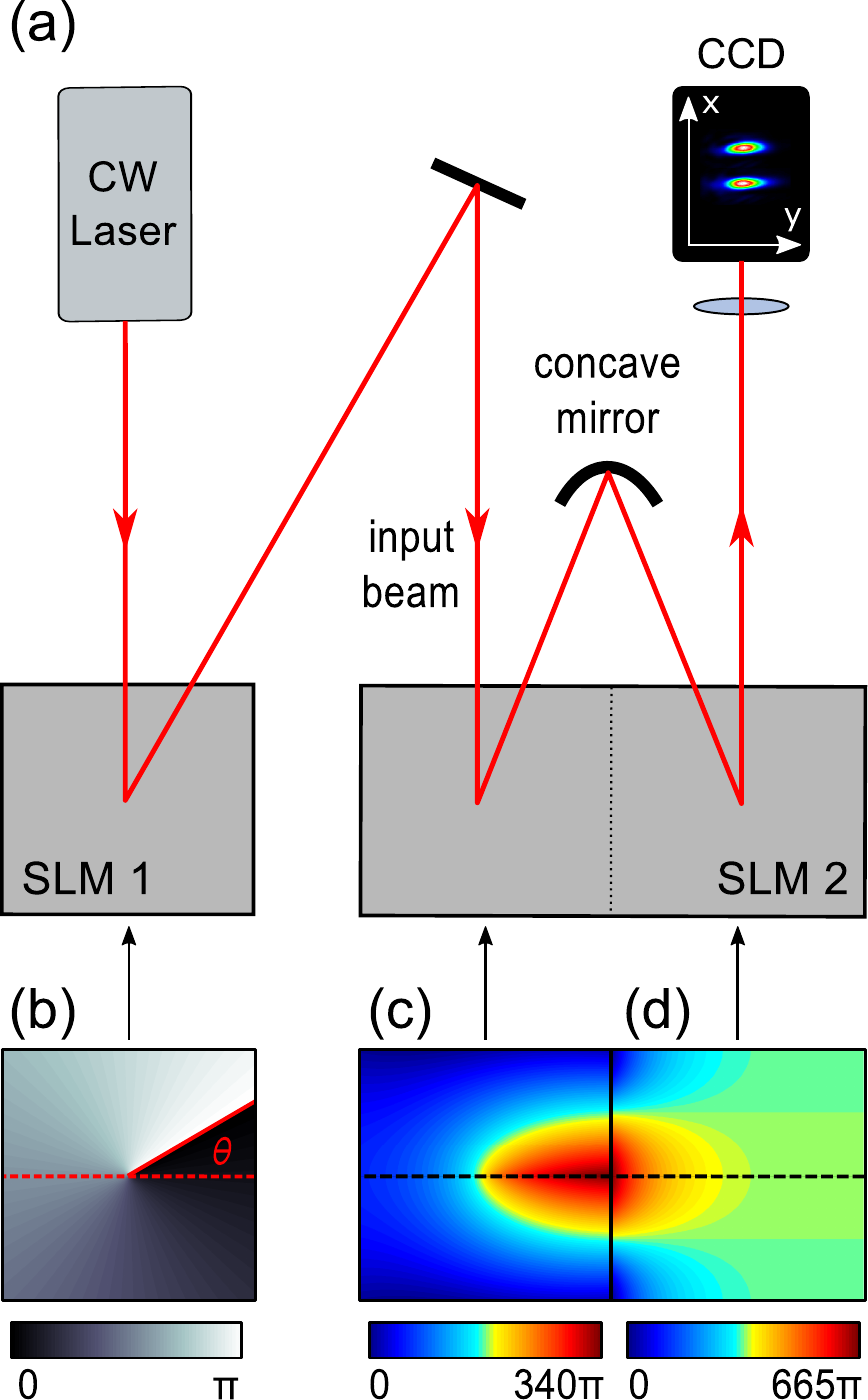}
	\caption{Experimental setup with the essential optical elements based on the setup by Berkhout et al. \cite{Berkhout_2010} (a). The phase patterns for generation of a light beam with OAM (b) are displayed on SLM 1 (pattern for $l = -0.5$ shown exemplary). The orientation angle $\theta$ can be adjusted with respect to the symmetry axis of the subsequent transformation (c) and phase correction pattern (d), which are displayed on SLM 2. The phase correction pattern (d) is flipped compared to an OAM sorting setup in transmission geometry to compensate for the mirroring due to the concave mirror. The effective focal length of the concave mirror is 100mm, the focal length of the lens for focusing onto the CCD camera is 400mm. The inset on the CCD in (a) shows a typical signal of two OAM modes sorted along the x-axis.
	}
	\label{fig:setup}
\end{figure}
\par
On the one hand, there are different methods for creating light beams carrying OAM. One simple method is to use an optical element which induces a vortex phase shift on the light beam wavefront, for example a phase plate\cite{Beijersbergen_1994} or a $l$-fold fork hologram, where the first order diffracted light beam carries the corresponding OAM state \cite{Heckenberg_1992, Bazhenov_1992}.
On the other hand, also the detection of OAM states can be realized in different ways. For example, phase elements for creation of optical vortices also can be used for detection of a single chosen OAM state by flattening the phase front and filtering the beam \cite{Leach_2002}. Also a cylindrical lens, which performs a 1D Fourier transformation of the light field, can be facilitated to measure the average OAM of a light beam \cite{Alperin_2016} or space-variant half-wave plates may be used\cite{Liu_2018}, which even allow to characterize vortex vector optical fields, but require sophisticated interferometric techniques.
\par
Here we provide an extension of OAM sorting, which enables simultaneous detection of multiple OAM states\cite{Berkhout_2010, Mirhosseini_2013}. This sorting technique is an imaging process with tailored phase elements which map the helical phase gradient of the OAM states onto a linear gradient. When focused by a lens, this geometry in turn results in output spots with sorted lateral positions depending on the topological charge of the OAM state. 
For purposes of data communication, it is customary to sort the intensity distribution at the output plane into discrete bins corresponding to integer OAM states, since it is known a priori that only a set of integer-valued OAM states needs to be identified. Instead, in spectroscopy of semiconductor systems such as vortices in exciton-polariton condensates whose topological charge directly carries over to the OAM of the emitted light field \cite{Lagoudakis2008,Dominici2018,Ma2017}, the OAM state of the sample signal is fully unknown in advance. Even fractional vortices with non-integer topological charge could appear. Thus, one must not make any prior assumption with respect to the OAM spectrum and even non-integer OAM values may occur. Identifying them requires a different approach of evaluating the output of the OAM sorting process. Here we demonstrate that calculating the centre of mass and sorted variance of the output intensity distribution similar to a series expansion is a reasonable approach. Within this framework, we study the imaging process of incoherent superpositions of integer OAM states and fractional OAM states with non-integer topological charge in detail. To this end, we define incoherent superpositions as states where the relative phase between the two integer OAM states is not fixed on average. We show how fractional OAM states can be identified. Furthermore, we identify the intrinsic and total OAM with the OAM sorting method. Finally we discuss the application of this method to vortices in exciton-polariton condensates, which are an ideal subject of research due to the direct optical accessibility. Our approach opens up the possibility to study different processes in exciton-polariton condensates such as vortex decays or vortex switching.

\section{Methods}
Our setup is shown in Fig. \ref{fig:setup}. We use a CW laser beam with a wavelength of 845\,nm, which is in the typical wavelength regime of exciton-polaritons in InGaAs microcavities. The polarization of the light is always linear, so that its average spin angular momentum is zero. Two spatial light modulators (SLMs) are facilitated for generation of OAM states and performing the OAM sorting process\cite{Berkhout_2010}. The specific parameters of the phase elements used in this setup are $a=(0.008\,\text{m})/(2\pi)$, $b=0.00477\,$m, $\lambda=845\,$nm and $f=0.095\,$m. The laser beam illuminates SLM 1 which displays a vortex phase pattern corresponding to the OAM state to imprint on the beam. For fractional OAM states, we also set the orientation angle $\theta$ of the phase discontinuity (see Fig. \ref{fig:setup}(b)) at this stage. The beam then is guided onto SLM 2, which is divided into two parts. One half displays a transformation pattern (see Fig. \ref{fig:setup}(c)), the other a phase correction pattern (see Fig. \ref{fig:setup}(d)).\\
\begin{figure}[h]
	\includegraphics[width=0.98\columnwidth]{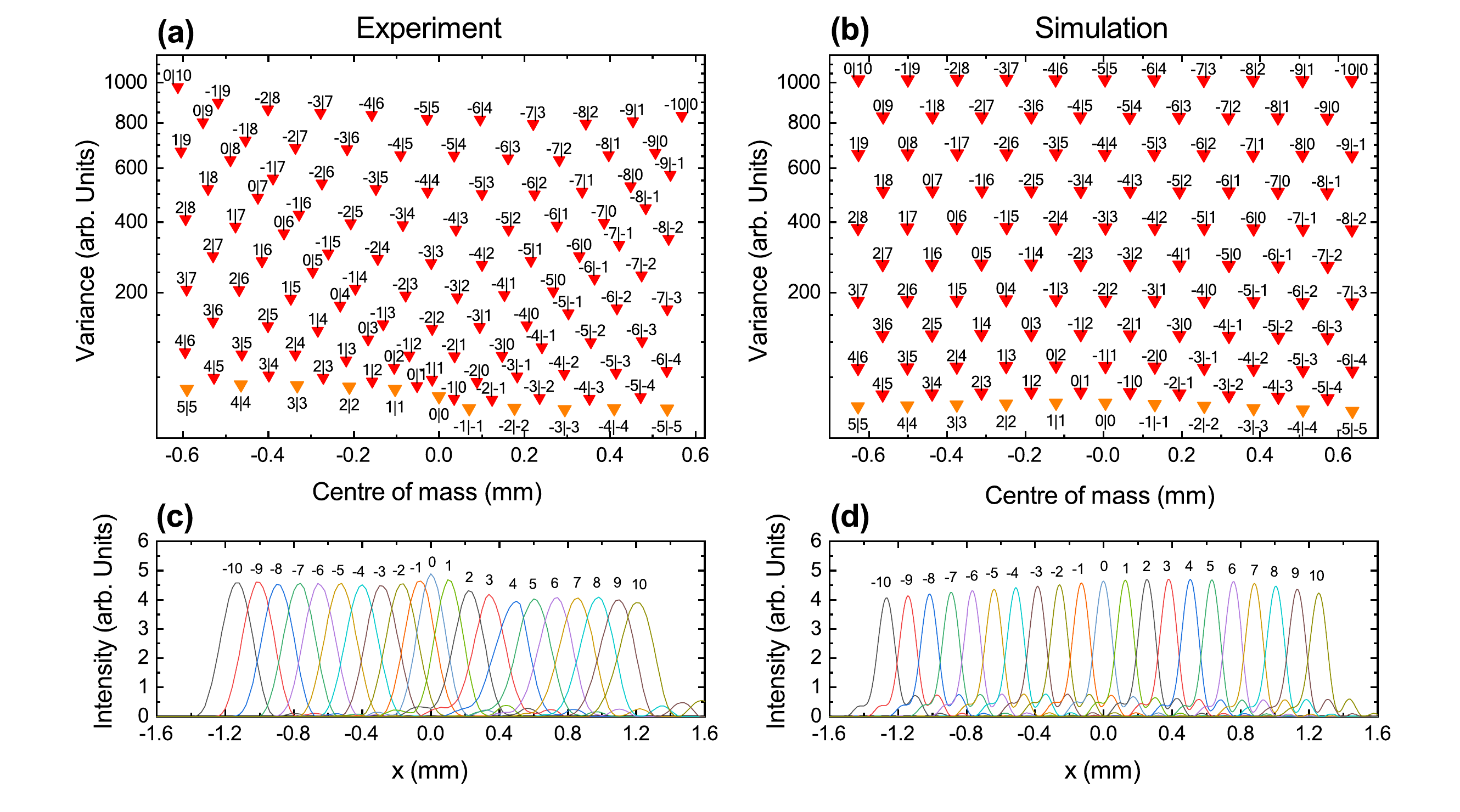}
	\caption{
		Incoherent superpositions of two integer OAM states with equal weight can be clearly identified by calculating the centre of mass and sorted variance both in experiment (a) and simulation (b). Red triangles show the centre of mass and variance for each superposition and the two numbers above the triangles identify the two integer OAM states forming the superposition. Orange triangles mark superpositions of two states with the same OAM, which effectively correspond to single OAM states. For comparison, the spatial profiles of single OAM states after sorting are plotted below for experiment (c) and simulation (d).
	}
	\label{fig:superposition_comp}
\end{figure}
The transformation pattern maps the helical phase gradient to a linear phase gradient. This pattern necessarily has some preferred axis. We place it such, that this preferred axis and the phase discontinuity at SLM 1 are aligned for $\theta$=0$^\circ$. The concave mirror then performs a Fourier transformation similarly to a lens and guides the beam onto the phase correction element. This pattern has the purpose to correct remaining deviations of the wavefront from the linear phase gradient and to reshape the beam to spots. From there, the resulting spots are focused by a lens onto the recording CCD camera, where the linear phase gradient results in a vertical deflection of the focused beam along the $x$ direction of the CCD camera. The magnitude of this deflection scales with the topological charge of the incident OAM state. For further evaluation, the spatial images of the output from the OAM sorting are integrated along the horizontal direction $y$, which is the the dimension perpendicular to the linear phase gradient axis. As the next step, in contrast to binning the output intensity distribution to integer OAM states, the centre of mass $x_{cms}$ and sorted variance $V$ of the OAM-sorted output intensity distribution are calculated as follows:
\begin{align}
	x_{cms}&=\sum_{i}x_i\cdot I(x_i),\label{eqn:xcm}\\
	V&=\sum_{i}(x_i-x_{cms})^2\cdot I(x_i),\label{eqn:variance}
\end{align}
where $i$ runs over all pixels, $x_i$ denotes the vertical position of pixel $i$ and $I(x_i)$ represents the intensity at pixel $i$. The centre of mass is now directly proportional to the average OAM value of the light field and the sorted variance is a measure for the width of the OAM spectrum, which provides information about the composition of the OAM states. We will now show that the sorted variance allows us to identify the components of superpositions of OAM states and also to distinguish superpositions of integer OAM states with fractional average OAM from true fractional OAM states. 

\section{Results}

A significant advantage of the evaluation of centre of mass and sorted variance by Eqs. (\ref{eqn:xcm}) and (\ref{eqn:variance}) is that single OAM states and incoherent superpositions of different states can be separated easily.
This is exemplarily shown in Fig. \ref{fig:superposition_comp} for light fields prepared as an equal mixture of two OAM states which are sorted afterwards. Figure \ref{fig:superposition_comp} also provides the results of simulations of the beam propagation and OAM sorting process using Matlab for comparison.
Each combination of two incoherently superposed states can be mapped onto a distinct combination of centre of mass position and sorted variance of the intensity distribution. This allows us to identify the two superimposed states just from the centre of mass and sorted variance both in simulation and experiment. Taking into account higher moments such as skewness and kurtosis, this principle can be extended to incoherent superpositions of more than two states or states with strongly differing intensities.

\begin{figure}
\centering
	\includegraphics[width=0.75\columnwidth]{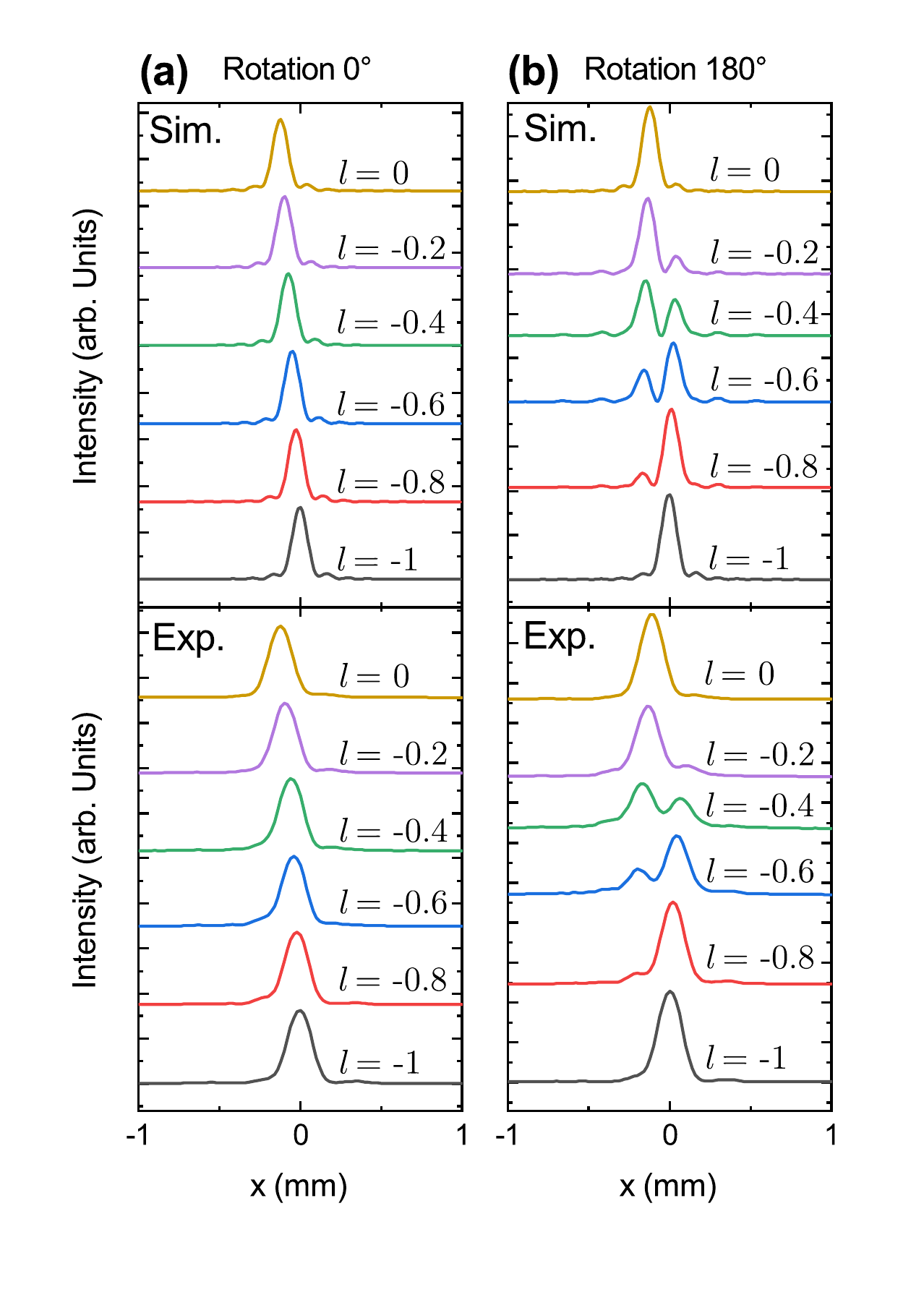}
	\caption{
		Spatial output profiles of sorted fractional OAM states from $l=-1$ to $l=0$ in experiment and corresponding simulations for orientation angles of $0^{\circ}$ (a) and $180^{\circ}$ (b). At an orientation of $180^{\circ}$, the peak of the sorted OAM signal splits for fractional OAM states.
	}
	\label{fig:fractional_sorting}
\end{figure}

However, the signal is not necessarily limited only to integer topological charges. Therefore, we also study fractional OAM states with non-integer topological charge. We find that in the OAM sorting process fractional OAM states result in quite different output intensity patterns depending on the orientation angle $\theta$ of the phase discontinuity of the fractional OAM state with respect to the symmetry axis of the phase elements for the OAM sorting process. Figure \ref{fig:fractional_sorting} shows the real space intensity pattern for sorting states of the same fractional OAM values, but different fractional OAM vortex orientations. For an orientation angle of $0^\circ$ the fractional OAM states are imaged onto single spots, which in principle shift linearly with OAM between the two nearest integer OAM states. When the orientation angle deviates from $0^\circ$, the peaks start to split into double peaks at positions, which do not match the peak positions for integer OAM states. At an orientation angle of $180^\circ$ the peak splitting becomes symmetric with respect to the central position between both nearest integer states.
\par
For a more detailed picture, we evaluate the centre of mass and sorted variance for a broader range of fractional OAM states at different orientation angles and compare them to the results of Alperin et al. \cite{Alperin_2017}. Here we focus on the relation between the intrinsic and extrinsic OAM components and the total OAM. The intrinsic OAM shows invariance with respect to rotations in the sense that a light field carrying only intrinsic OAM will result in the same OAM-sorted intensity distribution for any orientation angle $\theta$ between the direction of the phase discontinuity and the preferred axis of the transforming phase element. The intrinsic OAM component corresponds directly to the topological charge of the phase function used to generate the OAM beam and amounts exactly to $l$. However, it has been shown that the total OAM of the light beam may differ from this value. The difference is given by the extrinsic OAM component, which is equivalent to a net linear transverse momentum and usually associated with a change in the center of gravity of the beam. One of the simplest ways to perform such a shift and create extrinsic OAM lies in obstructing a part of the beam. For beams carrying extrinsic OAM components, the OAM-sorted intensity distribution will depend strongly on $\theta$, while for states with integer topological charge the extrinsic OAM always vanishes and the total OAM equals the rotation invariant intrinsic OAM. However, for fractional OAM states with non-integer topological charge, the non-rotational invariant extrinsic OAM component is mostly non-zero and the total OAM differs from the intrinsic one. The close link between fractional and extrinsic OAM has been explained as a consequence of the shape of the optical elements required to create fractional OAM states: These elements contain a steep phase discontinuity along some direction and this phase step allows nonpropagating evanescent waves to form, which are effectively equivalent to a local obstruction of the beam \cite{Alperin_2017}.

\begin{figure}
\centering
	\includegraphics[width=0.5\columnwidth]{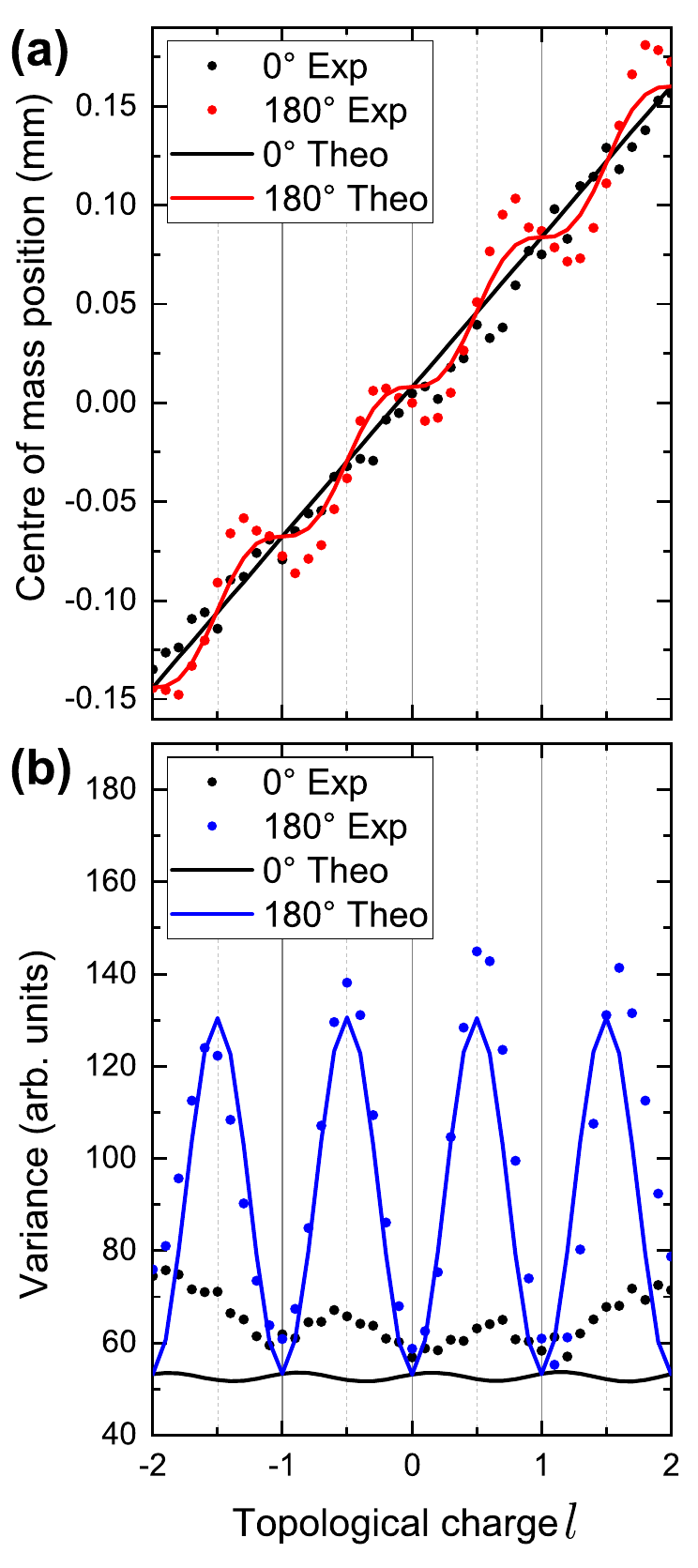}
	\caption{
	(a) Oscillations of the centre of mass values calculated from the intensity profiles of the respective OAM states with topological charges from \color{black}{$l=-2$ to $l=2$ for the two different orientation angles corresponding to intrinsic and total angular momentum in experiment and simulation. (b) The sorted variance strongly differs between both orientation angles in experiment and simulation due to the peak splitting as shown in Fig. \ref{fig:fractional_sorting}}.
	}
	\label{fig:cms_ozillations}
\end{figure} 

The results of our experiments and simulations for a wide range of $l$ and two particular orientation angles are shown in Fig. \ref{fig:cms_ozillations}. 
We find that in our setup using the OAM sorting process, the centre of mass which corresponds to the average OAM also shows a similar behaviour as the measurements of fractional OAM with a cylindrical lens from Alperin et al.\cite{Alperin_2017}. As discussed before, due to the phase discontinuity a fractional OAM state may show a non-zero total net linear momentum, which translates to the extrinsic OAM part of the beam. At the right orientation, this component vanishes and only the intrinsic OAM is measured. We find this orientation to be at the angle of $\theta=0^{\circ}$ in our setup. When the orientation is different, the measured average OAM shows oscillations. Under a certain angle the total angular momentum as sum of intrinsic and extrinsic OAM can be measured. We find this angle for the total OAM measurement to be $\theta=180^{\circ}$ in our setup by utilizing the fact that also half-integer OAM states have no extrinsic OAM components\cite{Alperin_2017}. Only for this orientation, the OAM-sorted intensity distribution shows the same centre of mass for half-integer OAM as the OAM-sorted intensity distribution for the intrinsic OAM measurement at $\theta=0^{\circ}$.
\par
The different behaviours with respect to both orientations leads to the insight that integer and non-integer OAM states can be distinguished already by comparing the centre of mass for the two orientation angles $\theta=0^{\circ}$ and $\theta=180^{\circ}$, except for the case of half-integer values. Here the sorted variance additionally needs to be taken into account to identify these states as fractional or integer OAM states. For half-integer OAM states the sorted variance shows significantly higher values at the orientation angle of $\theta=180^{\circ}$ compared to the orientation angle of $\theta=0^{\circ}$.
\par
It should be noted that there are obvious deviations between experiment and theory in Fig. \ref{fig:superposition_comp} as well as in Fig. \ref{fig:cms_ozillations}. In the latter case, the oscillations observed are even more pronounced than predicted by theory, which is a surprising result. Both effects can be traced back to using SLM 2 twice. The output of the transformation pattern shown on the left half of the SLM must be Fourier transformed and imaged onto the phase correction pattern on the right half of the SLM. This imaging process is subject to geometric constraints and effectively, one ends up either with clipping a part of the beam at the phase correction pattern or having a slightly tilted beam. In the present setup, the beam is slightly clipped at the phase correction pattern, which corresponds to removal of some high spatial frequency Fourier components from the beam. This frequency filtering results in the slight peak broadening seen in Fig. \ref{fig:fractional_sorting} and the deviations mentioned above. It can be avoided by adding a third SLM or phase element that can be moved independently of the other elements. Still, it is worthwhile to mention that due to the asymmetry of the sorted beam shapes, this filtering in the Fourier plane actually enhances the oscillations seen in Fig. \ref{fig:cms_ozillations}, which is an effect that might be useful for noisy signals.
\par
\begin{figure}
\centering 
	\includegraphics[width=0.97\columnwidth]{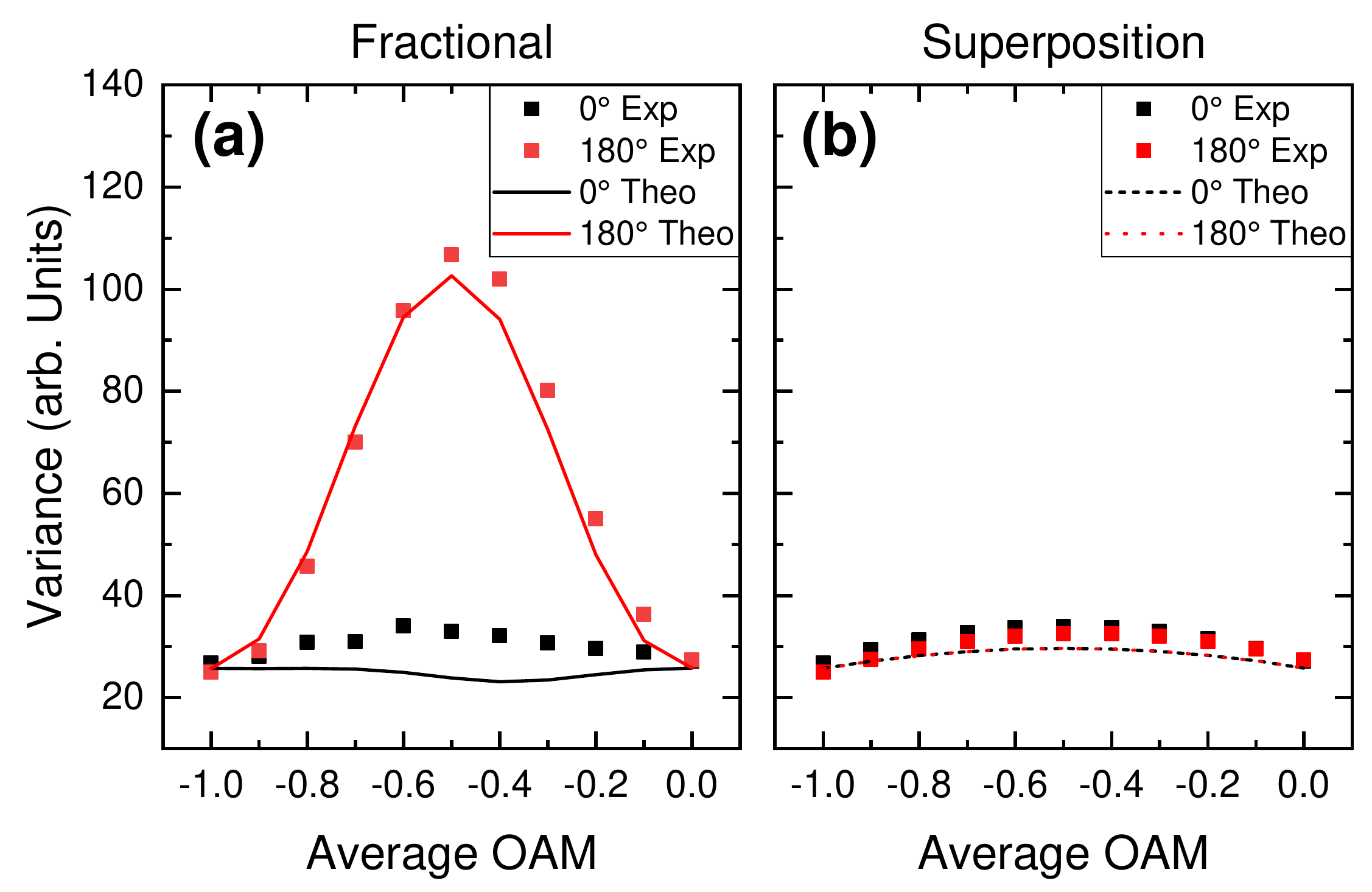}
	\caption{		
		Sorted variance of fractional OAM states (a) compared to incoherent superpositions of the integer OAM states $l=-1$ and $l=0$ with different weights chosen such that the beams carry the same average OAM as the fractional OAM states (b). The sorted variance for the fractional OAM states shows a much more pronounced dependence on the beam orientation than the incoherent superpositions.  
	}
	\label{fig:mixingvsfractional}
\end{figure}

Finally, we discuss whether it is also possible to distinguish fractional OAM states from incoherent superpositions with the same non-integer average OAM. In Fig. \ref{fig:mixingvsfractional}, we compare the sorted variance for fractional OAM states and incoherent superpositions of $l=-1$ and $l=0$ with their weights chosen such that the average OAM of the superposition also varies between -1 and 0. The sorted variance depends strongly on the relative orientation between the beam and the sorting element for fractional OAM states, but not for incoherent superpositions, as is shown exemplarily for the two orientation angles of $0^{\circ}$ and $180^{\circ}$. This difference can be understood by considering the underlying symmetry of the input signals. For a superposition state, the rotational symmetry of the underlying integer OAM states is retained while for a fractional OAM state this rotational symmetry is broken due to the presence of the phase discontinuity. When the sorting element converts the helical phase gradient to a linear phase gradient, the relative orientation of the beam with respect to the element determines the position on the linear phase gradient, where the steep phase step will be placed. For an orientation angle of $0^{\circ}$, one side of the step will be sorted to the beginning of the line, while the other side of the step will be sorted to the end of the line. Accordingly, the line will still show a linear phase gradient, which results in a small sorted variance when the beam is focused. For an orientation angle of $180^{\circ}$, the phase step will instead end up directly in the middle of the line, which causes the splitting of the line when the beam is focused and results in a large sorted variance.
\par
Therefore, the sorted variance of the output intensity distribution of the OAM sorting process does not change for a mixed state when changing the orientation angle while it does for a fractional OAM state. This enables us to experimentally distinguish between true fractional OAM states and mixed OAM states in an unknown sample signal by just rotating the signal with respect to the OAM sorting setup. Thus, the joint information gathered by recording the sorted signal at two orientation angles allows one to identify the OAM of the state much more easily compared to only evaluating the centre of mass in the $0^{\circ}$ orientation alone, which would require detectors with rather small pixel sizes.

\section{Discussion}
It is worthwhile to discuss the benefits and drawbacks of our approach to OAM sorting in comparison to other experimental techniques. It should be emphasized that we aim at applications in spectroscopy, especially semiconductor spectroscopy. As mentioned before, exciton-polaritons are a typical subject for OAM-resolved studies, so we will use them as an example system in the following discussion. For spectroscopic applications, it is highly important that the OAM sorting technique is compatible with other spectroscopy instruments. To this end, OAM sorting techniques that result in a signal sorted along one dimension are advantageous compared to other approaches because their output may directly be imaged onto the entrance slit of spectrometers or streak cameras, which will open up the possibility to record spectrally resolved polariton vortex spectra or time-resolved spectra with ps resolution.
\par
Especially with respect to investigations of vortex dynamics, it is natural to place strong emphasis on fractional OAM states and superpositions as these are the natural transient states between two stable vortex configurations. For ensemble-averaged measurements, it is not clear a priori, whether vortex decay will take place directly from one integer vortex state to the next at some randomly distributed time or whether this is a slow process taking place via a transient state of non-integer continuously changing vorticity. The former case will result in a superposition of OAM states being emitted from the sample, while the latter case will result in emission of a beam with fractional OAM. This topic is difficult to investigate using conventional phase interferometry as the phase patterns of integer and non-integer vortices look similar close to the vortex core and correctly identifying the outer vortex chain in the emission from real samples is a challenging task. Therefore, besides using OAM sorting based on continuous coordinates, especially the fact that our approach takes different orientation angles between the beam and the mode sorting element into account allows for an easier distinction between these two limiting cases compared to techniques relying on integer binning \cite{Berkhout_2011}. Instead, techniques that make use of integer binning are clearly much better suited for applications in quantum communication, where the task is to reliably estimate the OAM states of single photons and also for sorting beams with different spatial extension \cite{Lavery2013}. Here, it should also be emphasized that the sorting pattern provided by our setup depends on several geometric parameters such as beam size. Accordingly, the whole sorting process should be calibrated properly using reference states with well defined beam diameter and OAM.
\par
It should also be noted that the OAM superpositions that are of interest differ significantly between applications in quantum communication and spectroscopy. For quantum communication, OAM superpositions can be chosen freely and as has been shown \cite{Berkhout_2011}, integer binning allows one to identify superpositions of non-adjacent OAM states with excellent success rate. Superpositions of states with adjacent OAM are harder to identify, but can be avoided by completely avoiding usage of all odd or even OAM states. In spectroscopy, however, the opposite scenario is of interest. In most cases, a vortex decay will not transfer huge OAM values, but the initial and final state will differ by $\Delta l=\pm1$, so identifying superpositions of adjacent OAM states and distinguishing them from fractional OAM states is much more relevant for spectroscopy compared to quantum communication. The approach towards OAM sorting presented here is tailored towards such spectroscopic applications as it ensures one-dimensional OAM sorting required for performing further spectrally or time resolved studies and the ability to distinguish fractional OAM states from superpositions simultaneously.

\section{Conclusion}
We demonstrated that our setup is suitable for OAM spectroscopy and allows to identify unknown OAM states. Superpositions of multiple integer OAM states can be uniquely recognized using the centre of mass and sorted variance. Furthermore integer and non-integer states can be distinguished without any prior knowledge of the topological charge of the input signal. Also the orientation angles for sorting of intrinsic and total OAM of fractional OAM states are ascertained. It seems especially promising to apply this technique to excitations in semiconductor microcavities, such as exciton-polariton vortices, where the dynamics of vortices may be studied without the need for a phase reference on picosecond scales by replacing the CCD with a streak camera. Also, it seems attractive to miniaturize the OAM sorting setup for lab-on-a-chip devices using printable fiber micro-optics instead of SLMs \cite{Gissibl_2016,Weber_2017,Lightman_2017}.

\section*{Acknowledgements.} We gratefully acknowledge support from the DFG in the framework of TRR 142 within project A4.
\bibliography{Bibliography}

\end{document}